\documentclass[prl,twocolumn,showpacs]{revtex4-1}
\usepackage{graphicx}
\usepackage{times}
\usepackage{bm}
\usepackage{amsmath, verbatim}

\def\lsim{\mathrel{\rlap{\lower4pt\hbox{\hskip1pt$\sim$}}
    \raise1pt\hbox{$<$}}}                
\def\gsim{\mathrel{\rlap{\lower4pt\hbox{\hskip1pt$\sim$}}
    \raise1pt\hbox{$>$}}}                

\def\be{\begin{equation}}
\def\ee{\end{equation}}
\def\bea{\begin{eqnarray}}
\def\eea{\end{eqnarray}}
\def\bse{\begin{subequations}}
\def\ese{\end{subequations}}

\def\be{\begin{eqnarray}}
\def\ee{\end{eqnarray}}

\def\ga{{\ \lower-1.2pt\vbox{\hbox{\rlap{$>$}\lower5pt\vbox{\hbox{$\sim$}}}}\ }}
\def\la{{\ \lower-1.2pt\vbox{\hbox{\rlap{$<$}\lower5pt\vbox{\hbox{$\sim$}}}}\ }}

\def\beq{\begin{equation}}
\def\eeq{\end{equation}}

\def\bea{\begin{eqnarray}}
\def\eea{\end{eqnarray}}

\begin{document}

\title{To Close or Not to Close: The Fate of the Superconducting Gap Across the Topological Quantum Phase Transition in Majorana-Carrying Semiconductor Nanowires}

\author{Tudor D. Stanescu$^{1}$}
\author{Sumanta Tewari$^{2}$}
\author{Jay D. Sau$^3$}
\author{S. Das Sarma$^4$}

\affiliation{
$^1$Department of Physics, West Virginia University, Morgantown, WV 26506\\
$^2$Department of Physics and Astronomy, Clemson University, Clemson, SC
29634\\
$^3$ Department of Physics, Harvard University, Cambridge, MA 02138\\
$^4$Condensed Matter Theory Center and Joint Quantum Institute, Department of Physics, University of Maryland, College Park, Maryland, 20742-4111, USA}

\begin{abstract}
We investigate theoretically the low-energy physics of semiconductor Majorana wires in the vicinity of a magnetic field-driven topological quantum phase transition (TQPT). The local density of states at the end of the wire, which is directly related to the differential conductance in the limit of point-contact
tunneling, is calculated numerically.We find that the dependence of the end-of-wire local density of states on the magnetic field is nonuniversal and that the signatures associated with the closing of the superconducting gap at the Majorana TQPT are essentially invisible within a significant range of experimentally
relevant parameters. Our results provide a possible explanation for the recent observation of the apparent nonclosure of the gap at the Majorana TQPT in semiconductor nanowires.
\end{abstract}

\maketitle


The recently reported observation \cite{Mourik} of a zero bias peak (ZBP) in conductance measurements on semiconductor (SM) nanowires coupled to superconductors (SCs) may represent the first experimental evidence of Majorana fermions (MFs), which are theoretically predicted to exist in topological superconductors \cite{Sau}. Topological SC states capable of supporting MFs can be realized in SM wires with proximity-induced superconductivity by driving the system through a topological quantum phase transition (TQPT) using a suitably directed magnetic field  \cite{Long-PRB,Roman,Oreg}. At the TQPT, the SC gap necessarily vanishes  \cite{Sau,Long-PRB,Roman,Oreg, Read-Green}. The absence of any signature associated with the gap closure casts serious doubt on the Majorana fermion interpretation of the ZBP in the recent experiment \cite{Mourik}.

In this Letter, we offer a possible explanation for the observed non-closure of the gap. By solving numerically an effective tight-binding model for multiband nanowires with realistic parameters we show that, in the vicinity of the TQPT,  the amplitude of the low-energy states near the ends of the wire may be orders of magnitude smaller than the amplitudes of the localized MFs. Consequently, the contributions of these  states to the end-of-wire tunneling conductance and LDOS are essentially invisible, which results in an apparent non-closure of the gap in these quantities. By contrast, the closing of the gap mandated by the TQPT  is clearly revealed by other quantities, such as the total density of states (DOS) and the LDOS near the middle of the wire. We also show that this non-closure of the gap is non-universal, being dependent on the behavior of certain low-energy wave functions, and we identify the parameter regimes in which signatures associated with the closing of the gap are present in the end-of-wire LDOS.

 In Ref.~[\onlinecite{Sau}] Sau et al. proposed that a spin-orbit (SO) coupled semiconductor (SM) thin film with Zeeman spin splitting and proximity induced $s$-wave superconductivity could be used to realize Majorana fermions.
 For small Zeeman splitting $\Gamma$, the semiconductor is in a conventional (proximity-induced) superconducting  state with no MFs, while for $\Gamma$ larger than a critical value $\Gamma_c$ (corresponding to the TQPT where the SC gap vanishes), localized MFs \cite{Read-Green,Kitaev-1D,Sengupta-2001} exist at defects of the SC order parameter.  In subsequent works \cite{Long-PRB,Roman,Oreg,Annals,Alicea,Potter-Lee,Lutchyn-Stanescu,Zhang,Minigap} it was shown that in the 1D version of this system -- the so-called `semiconductor Majorana wire',  a direct physical realization of the Kitaev model \cite{Kitaev-1D} -- zero-energy MF states are trapped at the wire ends and protected from  regular fermionic excitations by a large mini-gap \cite{Long-PRB}
$\sim E_{qp} \sim 1$ K, where $E_{qp}$ is the proximity-induced bulk SC quasiparticle gap. The semiconductor Majorana wire, which has recently received considerable experimental attention \cite{Mourik,Deng,Rokhinson,Weizman,Harvard},  allows the detection of the zero-energy MF as a sharp zero bias peak in local charge tunneling measurements~\cite{Long-PRB,Sengupta-2001,R1} at experimentally realistic temperatures $T < E_{qp}$ \cite{Long-PRB}. Here, we study the signatures of  {\em nonzero} low-energy states in local end-of-wire measurements.

\begin{figure}[tbp]
\begin{center}
\includegraphics[width=0.48\textwidth]{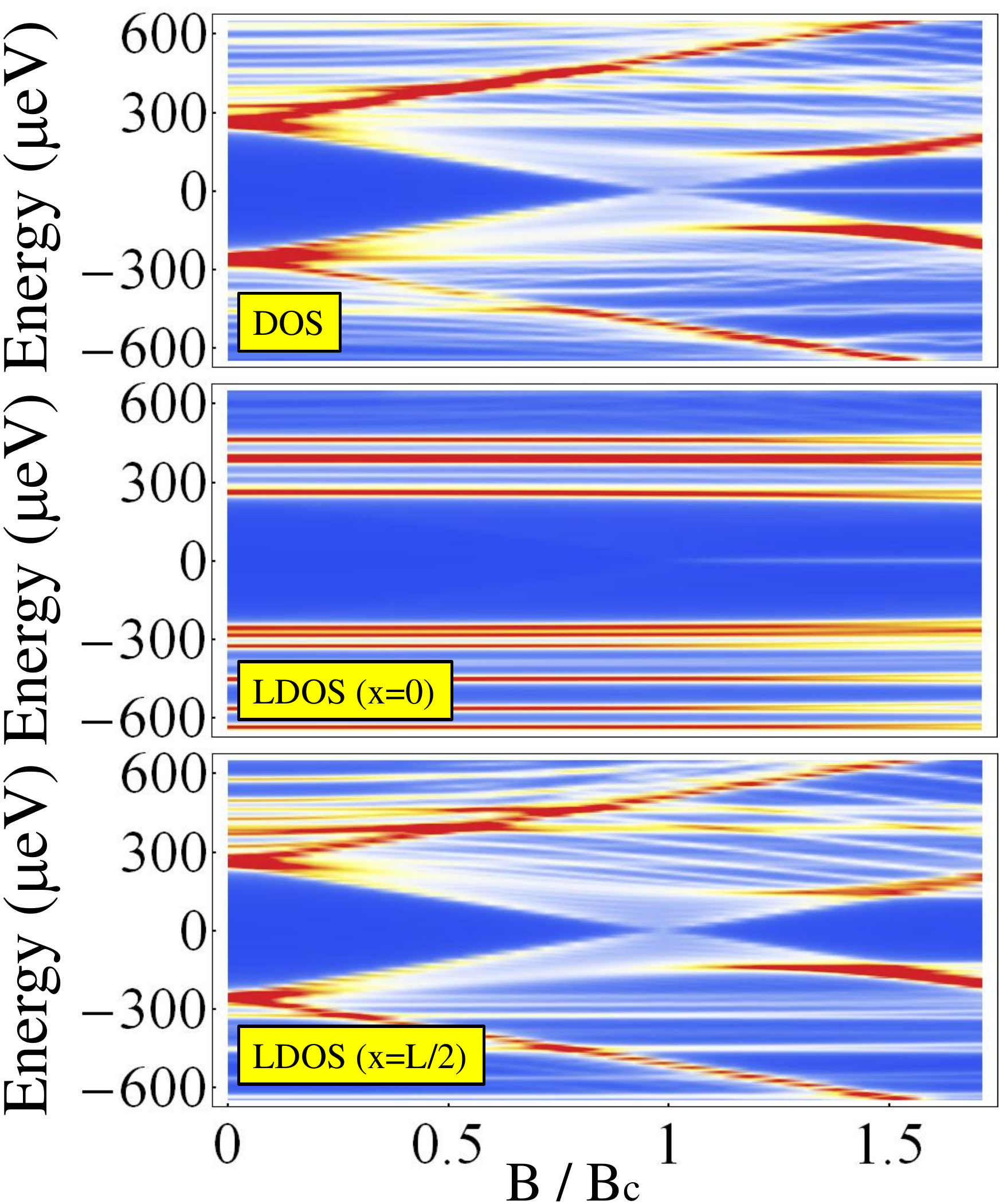}
\vspace{-7mm}
\end{center}
\caption{(Color online) {\em Top}: Total DOS as a function of $B$ for $\mu=\Delta/2$. The closing of the bulk gap is clearly  visible at the TQPT, $B=B_c\approx 0.2$ T. {\em Middle}: LDOS at the end of the wire as a function of $B$. The strong features associated with $\Delta \sim 250$ $\mu$eV are only weakly dependent on $B$. For $B>B_c$ a peak associated with the Majorana bound state is present at zero energy. The LDOS shows no visible signature of the bulk gap closing at the TQPT. {\em Bottom}: LDOS at the middle of the wire. Note the closure of the gap at the TQPT and the absence of the zero-energy Majorana peak. Traces for selected values of $B$ are provided in the Supplemental Material~\cite{suppl}.}
\vspace{-6mm}
\label{Fig1}
\end{figure}

We consider a rectangular SM nanowire with dimensions $L_x\gg L_y \sim L_z$ proximity coupled to an $s$-wave superconductor. A realistic model of the nanowire that includes the effects induced by proximity to the SC is solved numerically for a set of effective parameters  corresponding to InSb following the procedure described in Ref.~[\onlinecite{SLDS}] (see also the supplemental material for technical details~\cite{suppl}). The wire has a cross section $L_y\times L_z \approx 45$ nm $\times$ $50$ nm and is characterized by a Rashba spin-orbit coupling  $\alpha=0.2$ eV\AA. An external  Zeeman field $\Gamma= g^* \mu_B B/2$, where $B$ is the magnetic field and  $g^*=50$,  is applied  along the $x$-direction.
We couple the wire to an $s$-wave SC with a bulk gap $\Delta_0=1.5$ meV. The proximity-induced effective pair potential in the SM is $\Delta=0.25$ meV.
With increasing $\Gamma$, the wire evolves from a non-topological SC state with no MF to a topological SC state with  MFs localized near the ends via a TQPT at $\Gamma=\Gamma_c$. The SC  quasiparticle gap induced in the wire {\em must} vanish at the TQPT \cite{Long-PRB,Read-Green,SLDS}. Such closing of the bulk gap is  clearly visible in the total DOS,  as shown in the top panel of Fig. \ref{Fig1}.

The main finding of this Letter is that, despite the vanishing of the bulk SC gap at the TQPT, as mandated inflexibly by the theory \cite{Long-PRB,Read-Green,SLDS}, this gap closure may in fact \textit{not} be visible in charge conductance experiments that aim to probe the end-state MFs.
We show this by explicitly calculating the end-of-wire LDOS, which is qualitatively related to the charge current passed through the end of the nanowire coupled to a normal lead \cite{Mourik}. This relation becomes a close correspondence  in the limit of point contact tunneling \cite{Giamarchi}.
Since the LDOS can be related more directly to the physics at the microscopic level, we focus on this quantity to shed light on the recent experimental results.
We consider a 1D nanowire system with four occupied bands, i.e., four pairs of spin sub-bands, (see Fig.~2, top panel) and a chemical potential close to the minimum of the top band. 
This is similar to the experimental situation in Ref.~\onlinecite{Mourik}, where only a few  sub-bands are thought to be occupied (results for the single-band case are provided in the Supplemental Material~\cite{suppl}).
Specifically, we have $\mu=\Delta/2$, where the chemical potential is measured relative to the energy of the top occupied band at $k_x=0$ and $B=0$.
The results are shown in middle panel of Fig. \ref{Fig1}. A zero-energy peak is clearly visible above the critical magnetic  field $B_c$, as well as several strong features with energies $\geq \Delta=250 \mu$eV that depend  weakly on $B$. Note that there is no visible signature associated with the closing of the gap at the TQPT. This type of behavior is very similar to the experimentally observed dependence of the end-of-wire differential conductance on the magnetic field~\cite{Mourik}.   By contrast, the LDOS calculated at the middle of the wire (Fig. \ref{Fig1}, bottom) clearly shows the bulk gap closing but no Majorana fermion ZBP for $B > B_c$.

\begin{figure}[tbp]
\begin{center}
\includegraphics[width=0.48\textwidth]{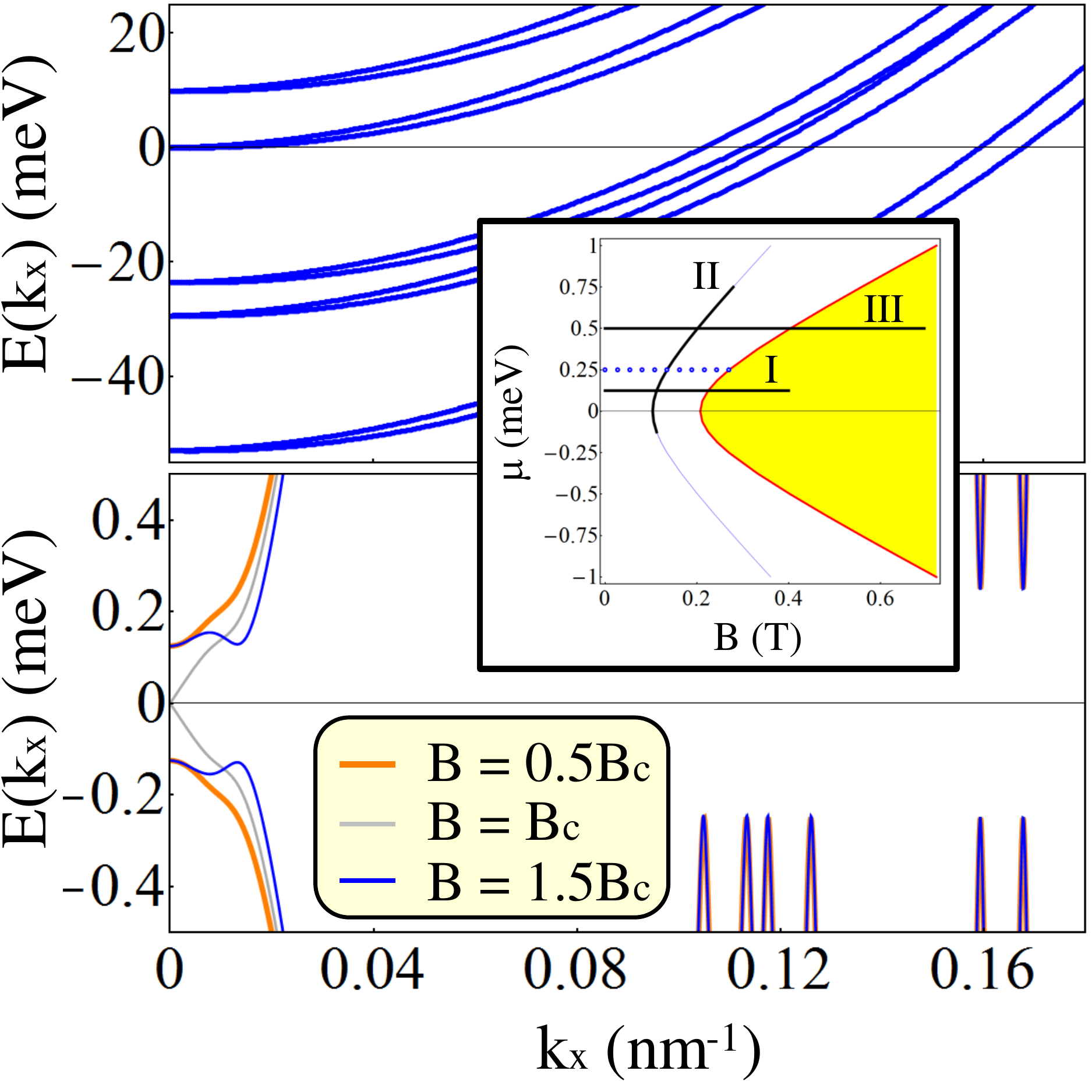}
\vspace{-8mm}
\end{center}
\caption{(Color online) {\em Top}: Spectrum of an infinite non--superconducting wire. The lowest three bands (six sub--bands) cross the chemical potential at Fermi wave vectors $k_F^{n\pm}>0.1$nm$^{-1}$,  while the top occupied band has $k_F^{4\pm}\ll 0.1$nm$^{-1}$. The fifth band in unoccupied.
{\em Bottom}: BdG spectrum for three different values of $B$. The induced SC pair potential is $\Delta=0.25$ meV. The three lower--energy bands depend weakly on $B$ and have nearly overlapping contributions characterized by six minima at energies $\approx 250$ $\mu$eV. By contrast, the top occupied band is strongly $B$--dependent (three distinct curves at $k_x<0.02$nm$^{-1}$) and is characterized by a gap that closes at the critical field $B_c\approx 0.2$ T.
{\em Inset}:  Relevant portion of the phase diagram showing the topological SC phase (yellow/light gray) and the topologically trivial phase (white). Fig. \ref{Fig1} corresponds to  a cut with $\mu=\Delta/2$  (path I), Fig. \ref{Fig4} is  for $\mu = 2\Delta$ (path III), while Fig. \ref{Fig5} corresponds to $B(\mu) = B_c(\mu)/2$ (path II).}
\vspace{-6mm}
\label{Fig2}
\end{figure}

To understand the somewhat unexpected behavior of the LDOS in Fig. \ref{Fig1},  we consider a nanowire with  a  chemical potential close to the bottom of the fourth band. This situation requires a low magnetic field to drive the system into the topological SC state, which is  consistent with the experimental conditions \cite{Mourik}. The normal state spectrum of the SM wire (for $B=0$) is shown in Fig. \ref{Fig2} (top panel).
The Bogoliubov-de Gennes (BdG) spectra associated with the four occupied bands are shown in Fig.~\ref{Fig2} (bottom panel) for different values of the magnetic field.
From the weak dependence on $B$ of the BdG minima associated with the three lower energy bands, it is clear that the prominent and weakly $B$-dependent finite energy features observed in the end-of-wire LDOS (Fig. \ref{Fig1}, middle panel) are associated with these bands. By contrast, the contribution to the BdG spectra associated with the top band
(or the `Majorana band') depends strongly on $B$. Note that the gap in the top band has a minimum at $k_x=0$ when $B\leq B_c$, vanishes at $B_c\approx 0.2$ meV, then reopens for $B>B_c$.
Thus, the low-energy physics in the vicinity of the TQPT and the gap closure at $B_c$ is controlled by the Majorana band,  while the bulk gaps at high Fermi momenta associated with the lower bands do not close at the TQPT and depend weakly on $B$.

Next, we consider a finite wire with  $L_x=4.5\mu$m and focus on the contributions to the LDOS coming from the BdG eigenstates associated with the lower bands.
In general, the contribution to the LDOS at the end of the wire coming from a given state $n$ depends on how fast the amplitude of the corresponding wave function $\Psi_n(x)$ increases as a function of $x$ (note that all wave functions must vanish at the wire ends).
For wires with confinement energy much larger than $\Delta$, states near the Fermi-level that are associated with the lower-bands
 have a large kinetic energy $\mu_{n}^{\rm eff}\gg \Delta, \Gamma$ (see Fig. \ref{Fig2} upper panel), which is the difference between the chemical potential $\mu$ and the bottom of the band $n$. These lower-band states have large Fermi wave-vectors and, consequently, are characterized by  amplitudes rapidly increasing away from the ends of the wire.
 We identify these  states as responsible for the weakly magnetic field dependent feature at $\approx 250$ $\mu$eV in Fig. \ref{Fig1}. Note that the lowest energies of the lower-band states have values close to the edge of the bulk gap, which disperses  weakly with $B$ and does not close at the TQPT (Fig.~\ref{Fig2}, bottom panel).

\begin{figure}[tbp]
\begin{center}
\includegraphics[width=0.48\textwidth]{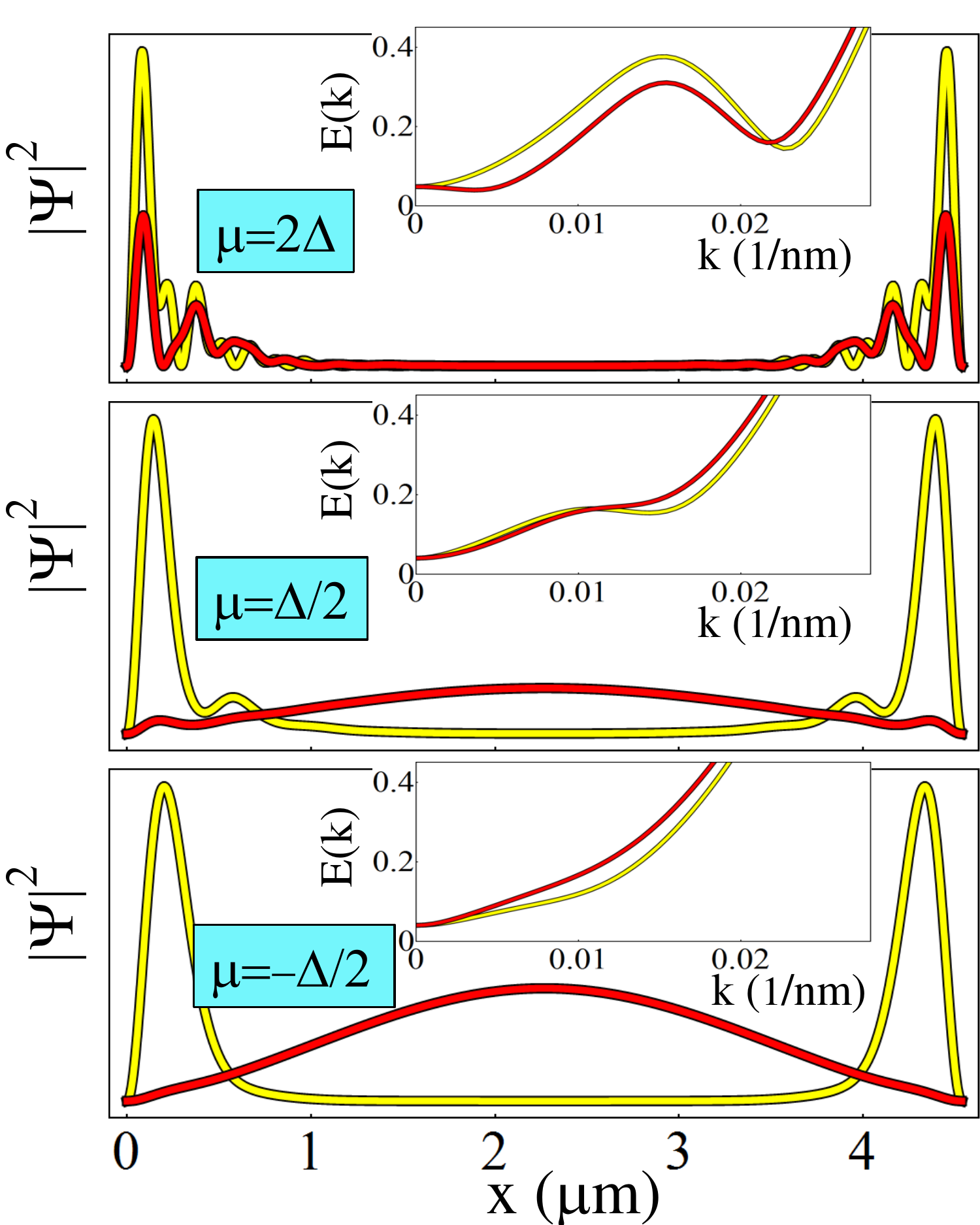}
\vspace{-7mm}
\end{center}
\caption{(Color online)  The lowest energy BdG states associated with the Majorana band near the TQPT for different values of $\mu$. The red (dark gray) lines correspond to $B=0.9B_c$, while the yellow (light gray) curves are for $B=1.1B_c$. The insets show the BdG spectra for an infinite wire near $k_x=0$. In the topological SC phase ($B>B_c$), the lowest-energy state is the MF state. In the non-topological SC phase ($B<B_c$), there is a crossover from the region with $\mu > \mu_c \approx \Delta$ characterized by localized lowest energy states (top panel) to the region with $\mu<\mu_c \approx \Delta$ where the states have mostly extended character (middle and bottom panels). The corresponding BdG spectra change from a structure with two minima ($\mu>\mu_c$) to a single minimum at $k_x=0$  ($\mu<\mu_c$).}
\vspace{-4mm}
\label{Fig3}
\end{figure}

By contrast, states associated  with the Majorana band have long characteristic wave-lengths and, consequently, their contributions to the end-of-wire LDOS are strongly suppressed whenever
these states are delocalized bulk states characterized by an envelope with
a vanishing amplitude near the ends. In general, the lowest energy states associated with the Majorana band also contain a localized component characterized by an envelope that decays exponentially away from the ends of the wire. In the non-topological SC phase, we find that the lowest energy states are delocalized (i.e., have a negligible localized component) when the chemical potential is below a certain crossover value $\mu_c(\Gamma)\sim{\cal O}(\Delta)$  (blue dots in the inset of Fig. \ref{Fig2}) and localized when $\mu > \mu_c$.
This delocalized-localized crossover is relatively sharp, being characterized by an energy scale of order $\Delta$. Also, we emphasize that there are multiple localized-delocalized crossovers characterized by values of the chemical potential $\mu_{c, n}$ slightly above the minimum of each band. The striking difference between the lowest energy states on the two sides of the crossover (for $n=4$) is illustrated in Fig. \ref{Fig3} (red/dark gray lines).
Note that in the topological SC phase ($B>B_c)$ the lowest energy state is always a Majorana bound state (yellow/light gray lines in Fig. \ref{Fig3}).

\begin{figure}[tbp]
\begin{center}
\includegraphics[width=0.48\textwidth]{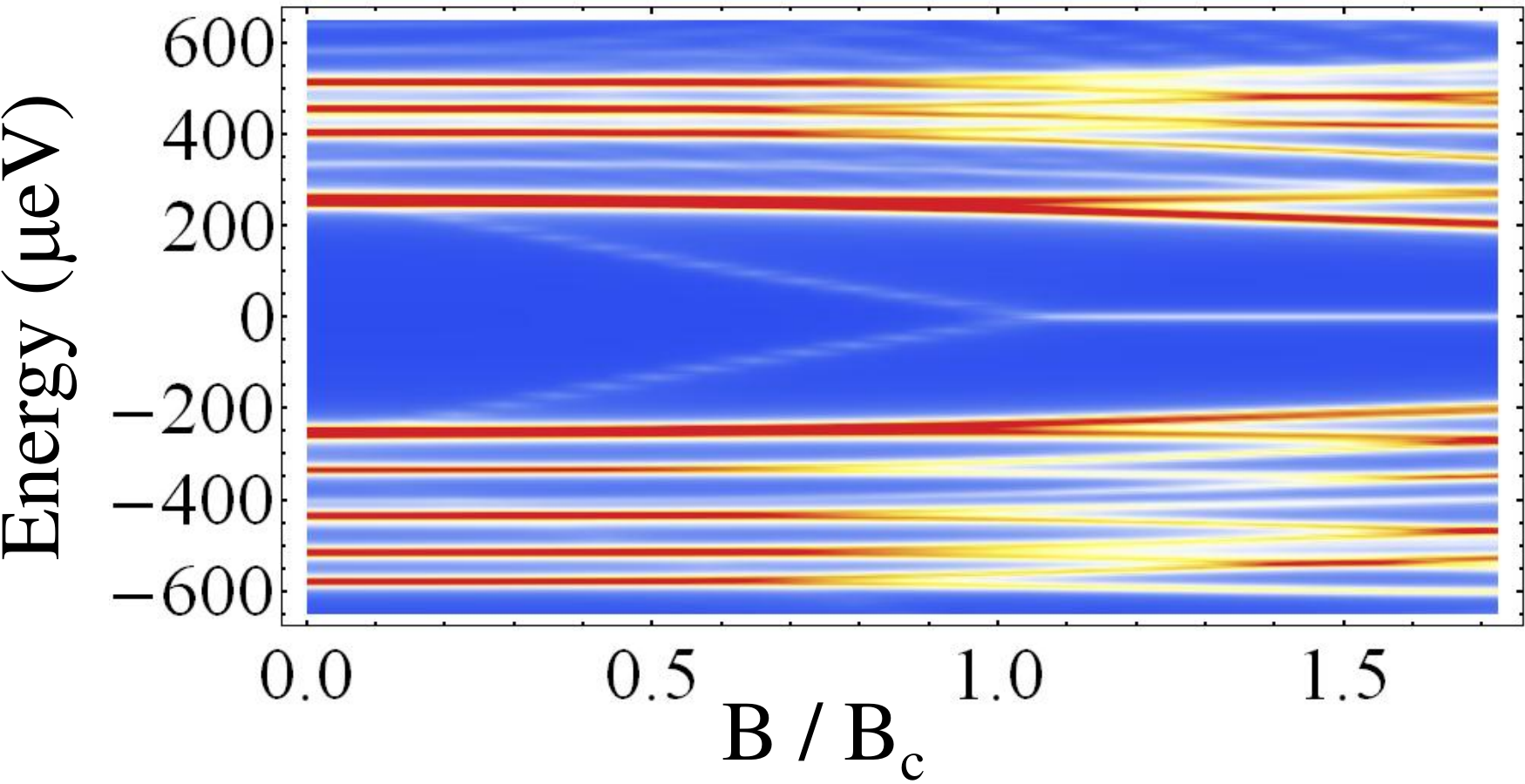}
\vspace{-10mm}
\end{center}
\caption{(Color online)  Dependence of the LDOS at the end of the wire on $B$ for $\mu = 2\Delta$. The gap-edge feature stemming from the localized end state clearly reveals the gap closing at the TQPT. For $B>B_c$ the finite energy states are de-localized and, consequently, there is no feature showing the re-opening of the gap. Traces for selected values of $B$ are provided in the Supplemental Material \cite{suppl}.}
\vspace{-4mm}
\label{Fig4}
\end{figure}

The localized/delocalized character of the low-energy states is directly reflected in the end-of-wire LDOS. If the system is in the delocalized regime, $\mu<\mu_c$, the signature of the low--energy states associated with the Majorana band is strongly suppressed. In particular, there is no visible signature of the gap closing at the TQPT. In this case, the dominant features originate from states in the lower bands, with energies $\geq \Delta$ that depend weakly on the magnetic field. Note that for $\mu=0$, which corresponds to the minimum of the critical Zeeman field $\Gamma_c(\mu) = \Delta$ (see the inset of Fig. \ref{Fig2}) and is probably very similar to the experimental conditions ~\cite{Mourik}, the system is in the delocalized region, hence no gap closure at the TQPT is visible in the end-of-wire LDOS. However, this behavior is not universal. Increasing the chemical potential above $\mu_c$ will move system into the a region characterized by localized lowest-energy states.  Since the energy of the localized states is at the bulk gap edge, the LDOS will contain a feature associated with the gap edge that will depend strongly on the magnetic field and will reveal the gap closure at the TQPT. This situation is illustrated in Fig. \ref{Fig4}.
We therefore predict that measurements probing other regimes of nanowire parameters may very well observe gap closing signatures, along with the emergence of MFs, depending on the details of the system.

\begin{figure}[tbp]
\begin{center}
\includegraphics[width=0.48\textwidth]{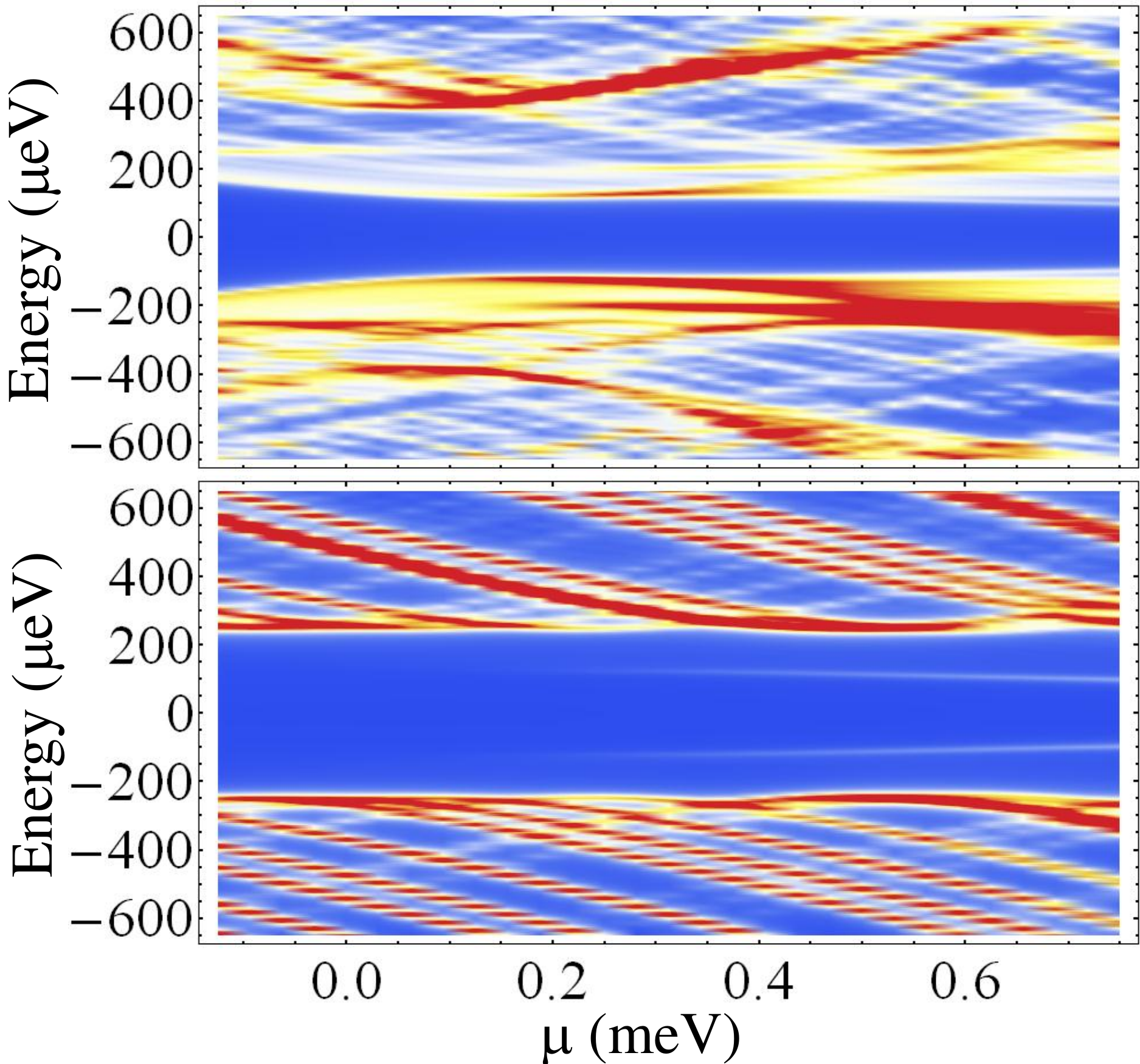}
\vspace{-5mm}
\end{center}
\caption{(Color online)  Total DOS (top) and LDOS at the end of the wire (bottom) as functions of $\mu$. The dependence on the chemical potential is along path II  (see inset of Fig. \ref{Fig2}). The gap is smaller than the pair potential $\Delta=250$ $\mu$eV, as shown by the DOS (top panel). In the LDOS, a feature associated with the gap edge becomes visible only in the localized region,  $\mu>\mu_c\approx 250$ $\mu$eV (blue dots in the inset of Fig. \ref{Fig2}).
}
\vspace{-4mm}
\label{Fig5}
\end{figure}

The qualitatively different behavior of the lowest energy BdG wave functions for $\mu < \mu_c \sim \Delta$ as compared to  $\mu > \mu_c$ can be correlated with the main features of the corresponding BdG spectra. As shown in the inset of Fig.~\ref{Fig3} (top panel) the BdG spectrum for $\mu > \mu_c$ has two well defined minima at two non-zero values of $k_x$.  This behavior is similar to that corresponding to the lower three bands (see Fig. \ref{Fig2}, lower panel).
By contrast, as shown in the insets of the middle and bottom panels in Fig. \ref{Fig3}, the BdG spectra for $\mu < \mu_c$ have only a single minimum at $k_x=0$.
In the cross-over region, $\mu \sim \mu_c$, the lowest energy BdG state for the finite wire is a linear superposition of localized and extended states.
Because of this clear pattern, we find that for $\mu\gtrsim \mu_c $ the end-of-wire LDOS generically reveals the dispersion of the bulk gap in the Majorana band via the localized end states at the gap edge, while for $\mu < \mu_c$ the end-of-wire LDOS does not disperse with $B$. This is further illustrated by the dependence of the LDOS on the chemical potential in Fig.~\ref{Fig5}. Note that the LDOS (bottom panel) shows features corresponding to the bulk gap edge only in the region $\mu > \mu_c$, while it shows no visible
signatures associated with the gap edge for $\mu < \mu_c$. We emphasize that this behavior is dictated by the localization properties of the low-energy wave functions in various regions of the phase diagram. These properties are not expected to change qualitatively in the presence of a smooth confining potential~\cite{Prada}, finite temperature~\cite{Lin}, or weak disorder~\cite{Disorder} (see the Supplemental Material for details \ref{suppl}).

We establish that the LDOS in the semiconductor Majorana wire is a non-universal quantity  sensitively dependent on the details of the system parameters and that it is entirely possible for the LDOS not to manifest any signature of the SC gap closing  through the TQPT.
When the chemical potential is in the vicinity of the minimum of an arbitrary band $n$,  we identify a sharp localized-delocalized crossover characterized by a value of the chemical potential (relative to the bottom of band $n$)  $\mu_c\sim\Delta$, where $\Delta$ is the proximity induced pair potential.  For $B < B_c$ and $\mu > \mu_c$,  the lowest energy BdG states associated with the top Majorana band are localized near the wire ends and contribute significantly to the end-of-wire LDOS. Since the energy of these localized states is close to the gap edge, the LDOS reveals the dispersion of the SC gap with $B$ and the eventual closing at the TQPT. By contrast, for $\mu < \mu_c$ the lowest energy BdG states associated with the Majorana band  decay near the wire ends. Consequently, their contribution to the end-of-wire LDOS is negligible and the  LDOS  does not reveal the bulk gap closing at the TQPT. This behavior is robust against disorder, if it is not too strong to destroy the topological phase.  In the presence of a disorder potential, the low-energy extended states will become localized inside random segments of the wire, but, typically, will still decay near the ends (see the supplementary section for details). Also, finite temperature will generate broadening effects, but will not shift the spectral weight.  
 Since the LDOS at the end of the wire is related to $dI/dV$, which is
the experimentally measured quantity, we expect that for $\mu < \mu_c$ the bulk gap closing at the TQPT should not be seen in  tunneling conductance measurements, although the signature of the MFs would clearly show up as a zero-bias-conductance peak. This is consistent with recent experiments \cite{Mourik}, which observe no feature associated with the bulk gap closing at $B=B_c$, yet a ZBP is visible for$B>B_c$. By contrast, charge conductance experiments in the regime $\mu > \mu_c$, or experiments that directly probe the bulk gap  \cite{Tewari-TQPT}, should reveal the dispersion of the bulk SC gap and its eventual closing at $B=B_c$.
Experiments probing the LDOS at the middle of the wire  would reveal signatures of the bulk gap closing through the TQPT, but no Majorana ZBP.

This work is supported by DARPA-MTO, NSF, DARPA-QuEST, JQI-NSF-PFC, Harvard Quantum Optics Center, and Microsoft-Q.

{\em Note Added}: After the original submission of this manuscript, a number of theoretical papers \cite{Pientka,Liu,Pikulin,Kells,Rainis,TS1,TS2,DSarma,Takei} appeared addressing various aspects of the experimental work on the possible observation of the Majorana zero mode in semiconductor--superconductor hybrid structures.


\vspace{1cm}

\begin{center}
{\Large Supplemental Material}
\end{center}

{\em Modeling}. We consider a semiconductor nanowire with rectangular cross section and dimensions $L_x \gg L_y\sim L_z$. The low-energy physics of the nanowire is described by the Hamiltonian 
\begin{eqnarray}
H_{\rm SM} &=& H_0 +H_{\rm SOI} = \sum_{{\bm i}, {\bm j}, \sigma} t_{{\bm i}{\bm j}}c_{{\bm i}\sigma}^{\dagger}c_{{\bm j}\sigma} -\mu \sum_{{\bm i}, \sigma} c_{{\bm i}\sigma}^{\dagger}c_{{\bm i}\sigma}  \nonumber \\
&+& \frac{i \alpha}{2}\sum_{{\bm i},{\bm \delta}}\left[ c_{{\bm i}+{\bm \delta}_x}^{\dagger}\hat{\sigma}_y c_{{\bm i}} -  c_{{\bm i}+{\bm \delta}_y}^{\dagger}\hat{\sigma}_x c_{{\bm i}} + {\rm h.c.} \right],  \label{Hsm}
\end{eqnarray}
where $H_0$, which includes the first two terms, describes nearest neighbor hopping on a simple cubic lattice with lattice constant $a$ with $t_{{\bm i}{\bm i}+{\bm \delta}} = -t_0$, where ${\bm \delta}$ are the nearest--neighbor position vectors.  In Eq. (\ref{Hsm}) the last term represents the Rashba spin-orbit interaction (SOI), $c_{{\bm i}}^{\dagger}$ is a spinor $c_{{\bm i}}^{\dagger}=( c_{{\bm i}\uparrow}^{\dagger}, c_{{\bm i}\downarrow}^{\dagger})$ with $c_{{\bm i}\sigma}^{\dagger}$ being the electron creation operators with spin $\sigma$, $\mu$ is the chemical potential,   $\alpha$ is the Rashba coupling constant, and $\hat{\bm \sigma}=(\sigma_x,\sigma_y,\sigma_z)$ are Pauli matrices. Since the number of degrees degrees of freedom in a finite wire is large (of the order $10^7$), yet Majorana physics is basically controlled by a reduced number of low-energy degrees of freedom  (of the order $10^3$--$10^4$), we project the problem into the low-energy subspace spanned by a certain number of low-energy eigenstates of $H_0$. Explicitly, the eigenstates of  $H_0$ are
\begin{equation}
\psi_{{\bm n}\sigma}({\bm i}) =\prod_{\lambda=1}^3 \sqrt{\frac{2}{N_\lambda+1}}\sin\frac{\pi n_\lambda i_\lambda}{N_\lambda+1} \chi_\sigma,    \label{psin}
 \end{equation}
 where ${\bm n}=(n_x, n_y, n_z)$ with $1\leq n_\lambda \leq N_\lambda$, and $\chi_\sigma$ is an eigenstate of the $\hat{\sigma}_z$ spin operator. The corresponding eigenvalues are
 \begin{equation}
 \epsilon_{\bm n} \!=\! -2 t_0 \left( \cos\frac{\pi n_x}{N_x\!+\!1} \!+\!  \cos\frac{\pi n_y}{N_y\!+\!1} \!+\!  \cos\frac{\pi n_z}{N_z\!+\!1}-3\right)\!-\!\mu_0, \label{epsn}
\end{equation}
where the chemical potential  $\mu_0$ is calculated from the bottom of the first band. The low-energy subspace is defined by the eigenstates satisfying the condition $\epsilon_{\bf n} <\epsilon_{\rm max}$, where the cutoff energy  $\epsilon_{\rm max}$ is of the order $75$--$100$meV. Using this low-energy basis, the matrix elements of the SOI Hamiltonian can be written explicitly as \begin{eqnarray}
&~&\langle\psi_{{\bm n}\sigma}|H_{\rm SOI}|\psi_{{\bm n^\prime}\sigma^\prime}\rangle = \alpha \delta_{n_z n_z^\prime}\left\{ \frac{1-(-1)^{n_x+n_x^\prime}}{N_x+1}(i\hat{\sigma}_y)_{\sigma \sigma^\prime} \right. \nonumber \\
&~&~~~~~~~~~\times \left. \frac{\sin\frac{\pi n_x}{N_x+1}\sin\frac{\pi n_x^\prime}{N_x+1}}{\cos\frac{\pi n_x}{N_x+1}-\cos\frac{\pi n_x^\prime}{N_x+1}}\delta_{n_y n_y^\prime} - [x \Leftrightarrow y] \right\}, \label{HSOI}
\end{eqnarray}
where the second term in the parentheses is obtained from the first term by exchanging the  $x$ and $y$ indices. Note that the SOI Hamiltonian has the structure $H_{\rm SOI} = H_{\rm SOI}^x +H_{\rm SOI}^y$, where the first term represents the intra-band Rashba coupling, while $H_{\rm SOI}^y$ couples bands with different $n_y$ indices. 
\begin{figure}[tbp]
\begin{center}
\includegraphics[width=0.48\textwidth]{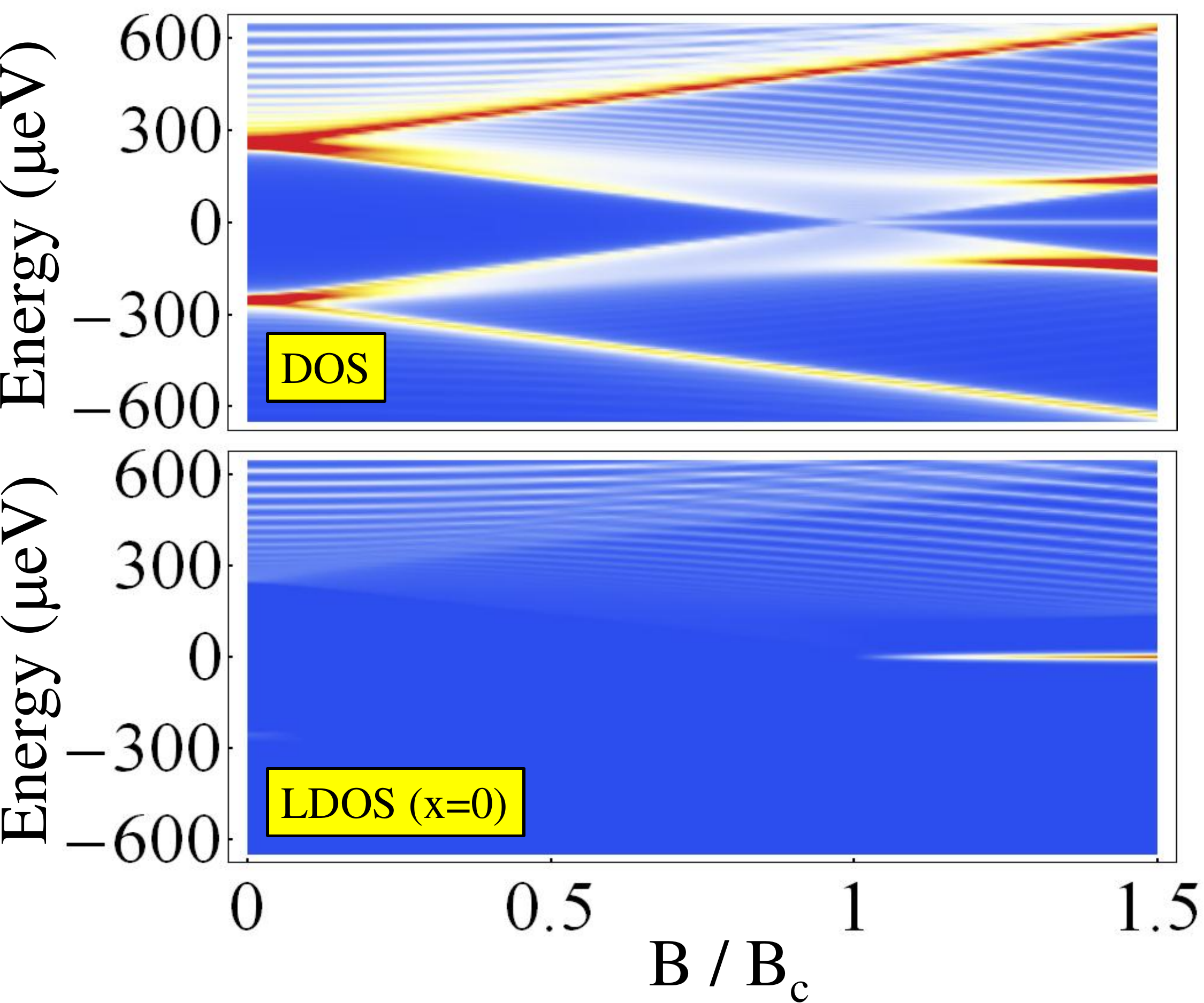}
\vspace{-7mm}
\end{center}
\caption{(Color online) {\em Top}: Total DOS as a function of the magnetic field for a system with single band occupancy. For $B=0$, the chemical potential is set at the bottom of the first band, i.e., $\mu=0$ and $n_{\rm top}=1$. The closing of the bulk gap is clearly  visible at the TQPT, $B=B_c\approx 0.2$ T. {\em Bottom}: LDOS at the end of a single channel nanowire as a function of $B$.  For $B>B_c$ a peak associated with the Majorana bound state is present at zero energy. The LDOS shows no visible signature of the bulk gap closing at the TQPT. Note the absence of any strong feature associated with the induced gap $\Delta \sim 250$ $\mu$eV.}
\vspace{-6mm}
\label{sFig1}
\end{figure}
In addition to SOI, a critical ingredient for realizing Majorana fermions in semiconductor nanowires is represented by the Zeeman field. We consider that the Zeeman splitting $\Gamma$ is generated by applying a magnetic field oriented along the wire (i.e., along the $x$-axis),  $\Gamma = g \mu_B B_x /2$. The corresponding matrix element in the low-energy basis are 
\begin{equation}
\langle\psi_{{\bm n}\sigma}|H_{\rm Zeeman}|\psi_{{\bm n^\prime}\sigma^\prime}\rangle =\Gamma \delta_{{\bm n}{\bm n}^\prime} \delta_{\bar{\sigma} \sigma^\prime}, \label{HZeemannn}
\end{equation}
where $\bar{\sigma}=-\sigma$.

The third key ingredient is the proximity-induced superconductivity (SC).  After integrating out the SC degrees of freedom, the SC proximity effect is described by a self-energy term that has the form~\cite{SLDS} 
\begin{align}\label{eq:Sigma_clean}
\Sigma(\omega)=-\gamma\left[ \frac{\omega + \Delta_0 \sigma_y\tau_y}{\sqrt{ \Delta_0^2-\omega^2}}+\zeta \tau_z\right],
\end{align}
where $\gamma$ is the effective SM-SC coupling, $\tau_x$ and $\tau_z$ are Pauli matrices in the Nambu space, $\Delta_0=1$meV is the pair potential of the bulk SC, and $\zeta$ is a proximity-induced shift of the chemical potential. In the present calculations we take $\zeta=0$. As shown in detail in Ref. \onlinecite{SLDS}, within the static approximation $\sqrt{ \Delta_0^2-\omega^2} \rightarrow \Delta_0$, the self-energy becomes $\Sigma(\omega)\approx -\gamma/\Delta_0 \omega - \gamma \sigma_y\tau_y$
and the low-energy physics of the SM nanowire with proximity-induced SC can be described by an effective Bogoliubov-de Gennes Hamiltonian. 
\begin{figure}[tbp]
\begin{center}
\includegraphics[width=0.48\textwidth]{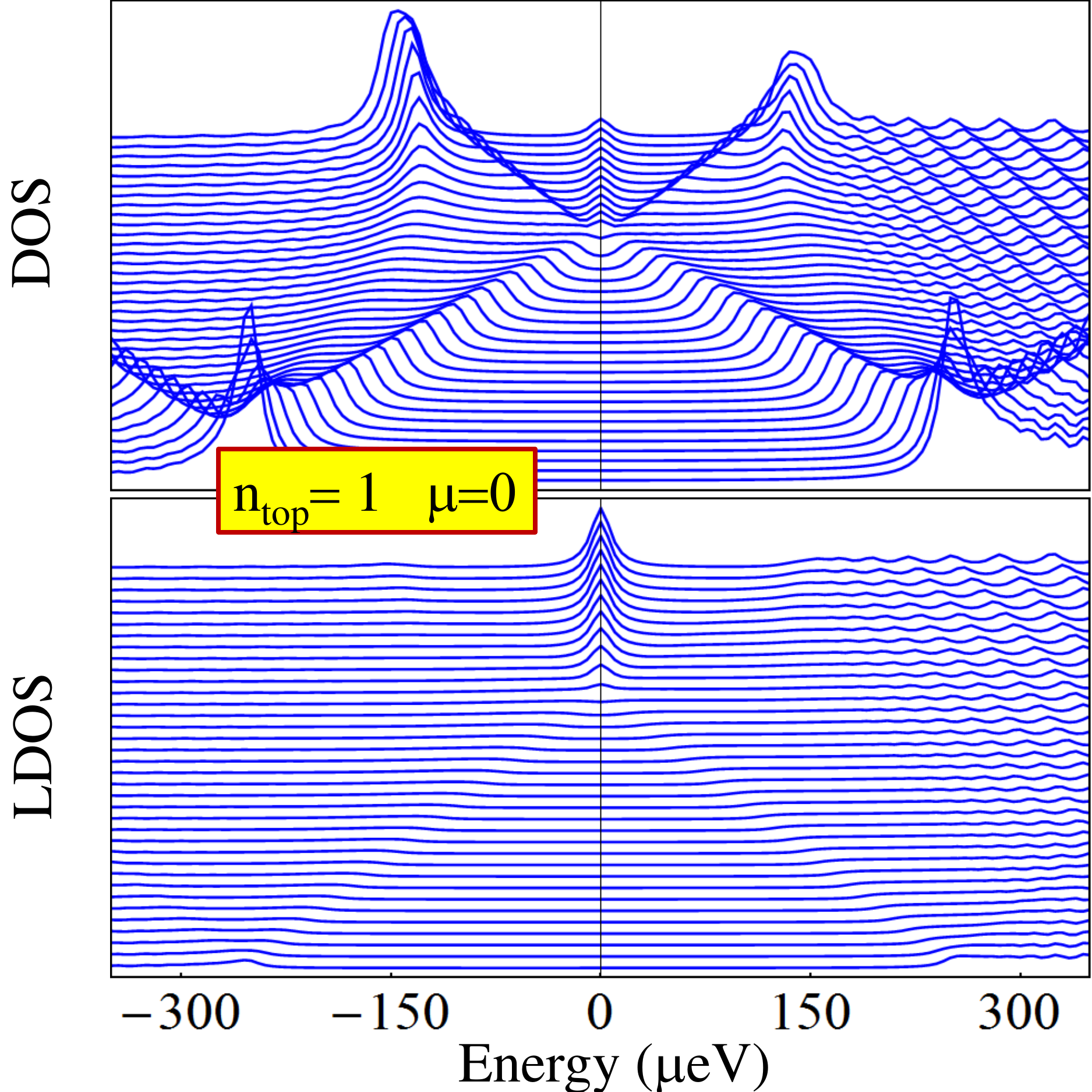}
\vspace{-8mm}
\end{center}
\caption{(Color online) Density of states $\rho(\omega)$ (top) and end-of-wire local density of states $\widetilde{\rho}_{\rm end}(\omega)$ (bottom) for a system with single band occupancy.  The traces (offset for clarity) represent constant magnetic field cuts through the color scale diagrams shown in Fig. \ref{sFig1}. The DOS and LDOS are calculated using Eqns. (\ref{dos}) and (\ref{ldos1}-\ref{ldos2}), respectively.}
\vspace{-4mm}
\label{sFig2}
\end{figure}
This approximation is valid, strictly speaking, at energies much lower than $\Delta_0$, but represents a very good approximation even for $E\sim\Delta_0/2$. Explicitly, the matrix elements of the effective BdG Hamiltonian can be written as
\begin{eqnarray}
 H_{\rm BdG}({\bm n},{\bm n}^\prime) &=& Z \left[\epsilon_{\bm n} \delta_{{\bm n n}^\prime} + \Gamma \sigma_x \delta_{{\bm n}{\bm n}^\prime} + \langle H_{\rm SOI}^x\rangle_{{\bm n n}^\prime} \right]\tau_z \nonumber \\
&+& Z \langle H_{\rm SOI}^y\rangle_{{\bm n n}^\prime} + \Delta \sigma_y \tau_y,  \label{Hbdg}
\end{eqnarray} 
where $\epsilon_{\bm n}$ is given by Eq (\ref{epsn}) and the SOI matrix elements are given by (\ref{HSOI}). Note that energy scale for the SM nanowire is renormalized by a factor $Z= (1+\gamma/\Delta_0)^{-1}$ due to the SC proximity effect. This renormalization is determined by the term in the self-energy (\ref{eq:Sigma_clean}) that is proportional to $\omega$ (in the static approximation). The pairing term in Eq. (\ref{Hbdg}) is derived from the corresponding contribution to the self-energy (\ref{eq:Sigma_clean}) and is proportional to the induced pair potential $\Delta=\gamma\Delta_0/(\gamma+\Delta_0)=250\mu$eV. The effective model described by Eq. (\ref{Hbdg}) is solved numerically 

\begin{figure}[tbp]
\begin{center}
\includegraphics[width=0.48\textwidth]{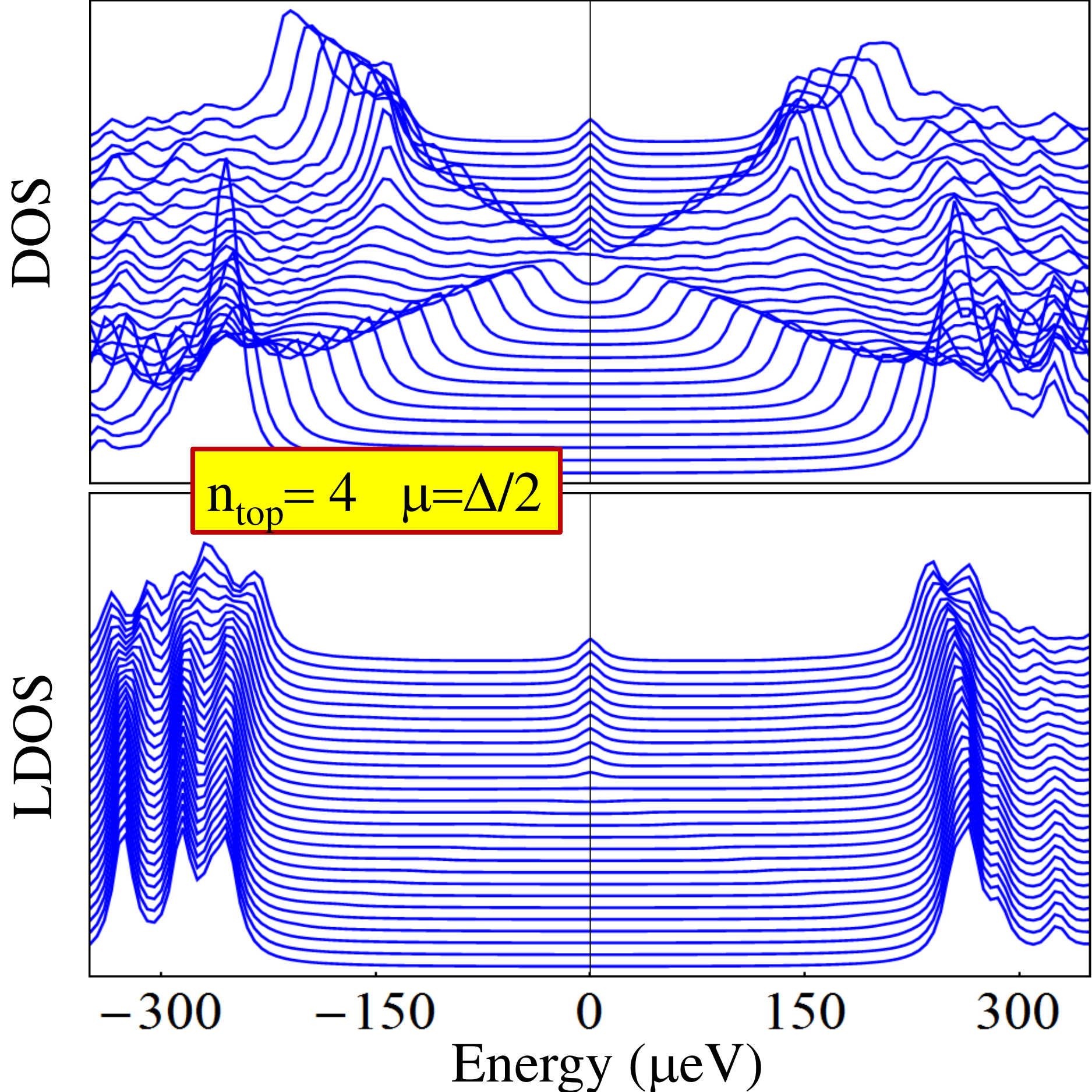}
\vspace{-7mm}
\end{center}
\caption{(Color online) Magnetic field traces corresponding to Fig. 1 (top and middle panels) from the main text. The dominant features above the induced gap $\Delta=250\mu$eV are due to states from the low-energy bands. Note that these features are absent in the single-band case (see Fig. \ref{sFig2}). The chemical potential corresponds to the ``delocalized'' regime, hence in the LDOS the are no signatures associated with the closing of the gap at the critical point. }
\vspace{-4mm}
\label{sFig3}
\end{figure}

{\em Density of states (DOS) and local density of states (LDOS)}. 
Diagonalizing the BdG Hamiltonian generates the eigenstates $\Phi_\nu=(u_\nu, v_\nu)$ and the corresponding energies $E_\nu$ with $E_{-\nu}=-E_{\nu}$, where $\nu = \pm 1, \pm 2, \dots$ and  $E_1 \leq E_2 \leq E-3\leq \dots$. The components   
$u_\nu$ and $v_\nu$ on the Nambu spinor  represent the particle and hole contributions, respectively. The density of states is defined as 
\begin{equation} 
\rho(\omega) = -\frac{1}{\pi} \sum_{\nu\leq 1} {\rm Im}\left[\frac{|u_\nu|^2}{\omega - E_\nu +i\eta} + \frac{|v_\nu|^2}{\omega + E_\nu +i\eta}    \right], \label{dos}
\end{equation}
where $|u_\nu|^2 = \sum_{{\bm n}, \sigma} u_\nu^*({\bm n}, \sigma) u_\nu({\bm n}, \sigma)$ and a similar expression holds for $|v_\nu|^2$. To calculate the LDOS, we define the real-space functions $\widetilde{u}_\nu({\bm i}, \sigma) = \sum_{\bm n} \psi_{{\bm n}, \sigma}({\bm i}) u_\nu({\bm n}, \sigma)$ and , similarly, $\widetilde{v}_\nu({\bm i}, \sigma)$.
Using these functions, the LDOS can be expressed as 
\begin{equation} 
\widetilde{\rho}({\bm i}, \omega) = -\frac{1}{\pi} \sum_{\nu\leq 1, \sigma} {\rm Im}\left[\frac{|\widetilde{u}_\nu({\bm i}, \sigma)|^2}{\omega - E_\nu +i\eta} + \frac{|\widetilde{v}_\nu({\bm i}, \sigma)|^2}{\omega + E_\nu +i\eta}    \right]. \label{ldos1}
\end{equation}
Note that $\sum_{\bm i} \widetilde{\rho}({\bm i}, \omega) =\rho(\omega)$. We note that in this work we are not interested in the LDOS at a specific point in the wire defined on the atomic scale, but rather in the average LDOS near the end of the nanowire, which is more closely related to the experimentally measured differential tunneling conductance. Consequently we define the end-of-wire (average) LDOS as 
\begin{equation}
\widetilde{\rho}_{\rm end}(\omega) = C \sum_{i_z, i_y}\sum_{i_x\leq i_{\rm end}} \widetilde{\rho}({\bm i}, \omega), \label{ldos2}
\end{equation}
where $C$ is a normalization constant and $i_{\rm end}=l_{\rm end}/a$, with $a$ being the lattice constant,  defines the characteristic length scale of the end-of-wire region. In the calculations  $l_{\rm end}\approx 20$nm, but the results do not change qualitatively if this quantity is varied in the range $1$-$10^2$nm.  

\begin{figure}[tbp]
\begin{center}
\includegraphics[width=0.48\textwidth]{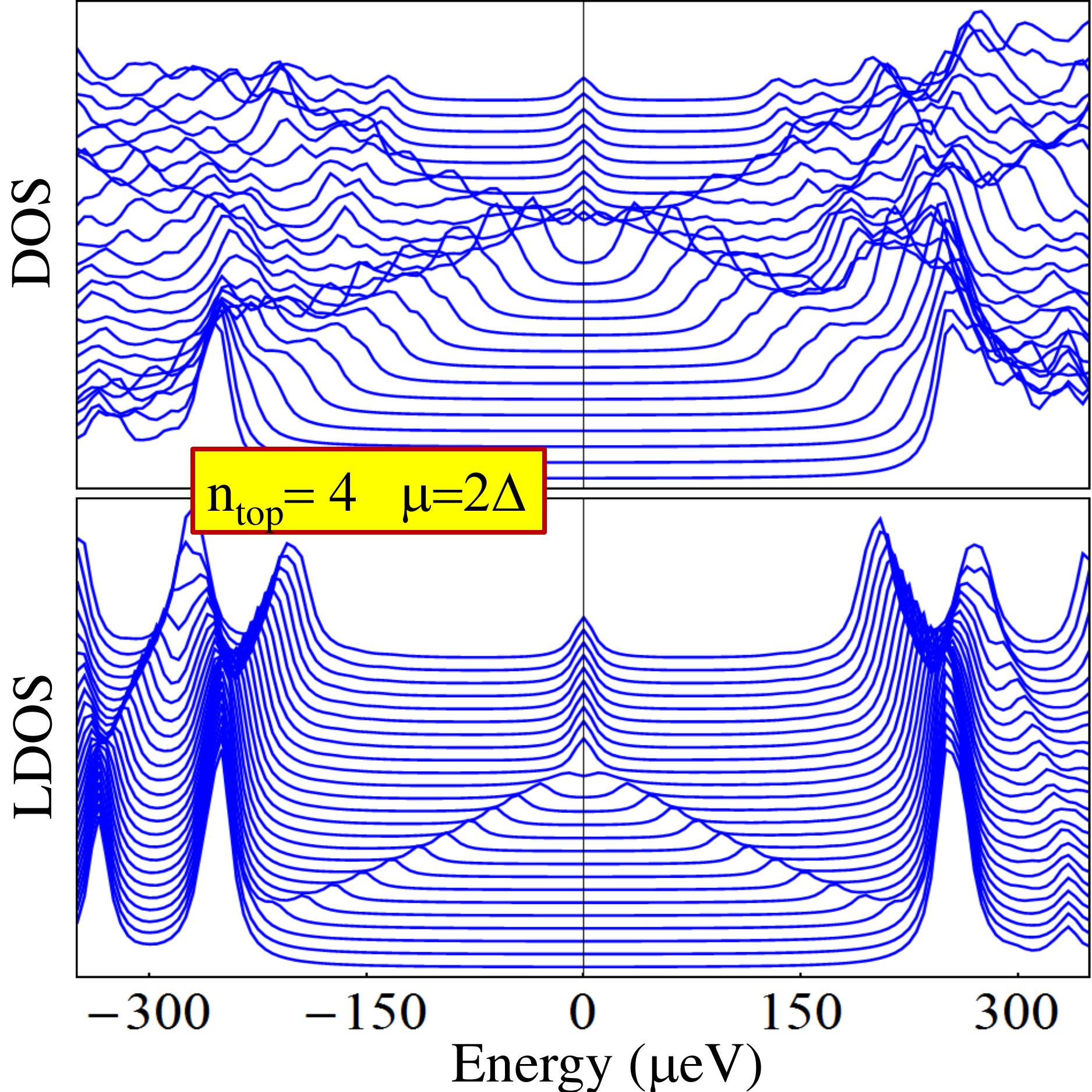}
\vspace{-10mm}
\end{center}
\caption{(Color online)  Magnetic field traces in the ``localized'' regime. The bottom panel represents the same quantity as Fig. 4 from the main text. The peaks that reveal the closing of the gap at the TQPT are generated by states from the top band that are localized near the end of the wire.}
\vspace{-4mm}
\label{sFig4}
\end{figure}

{\em The single-band case and the relative magnitude of various features in the DOS and LDOS}. 
Our main findings are: i) In a Majorana nanowire with a chemical potential close to the bottom of a band (i.e., in the ``delocalized'' regime)  the end-of-wire LDOS  does not manifest any signature of the gap closing though the TQPT. ii) In a multi-band Majorana nanowire, the end-of-wire LDOS exhibits strong features at an energy scale associated with the induced SC gap $\Delta$. These features have a weak magnetic field dependence and are generated by states from the low-energy bands. In the case of single-band occupancy, $n_{\rm top}=1$, property i) still holds, while  ii) is not applicable, as there is no other band with lower energy. To illustrate this case, we calculate the DOS and end-of-wire LDOS for a system with one occupied band and $\mu=0$. The results are shown in Fig. \ref{sFig1}. Notice that the Majorana peak can be clearly seen in the LDOS, but there are no features associated with the closing of the gap. Also, in contrast with the multi-band case (see Fig. 1 in the main text), there are no features associated with the induced gap $\Delta$. 

\begin{figure}[tbp]
\begin{center}
\includegraphics[width=0.48\textwidth]{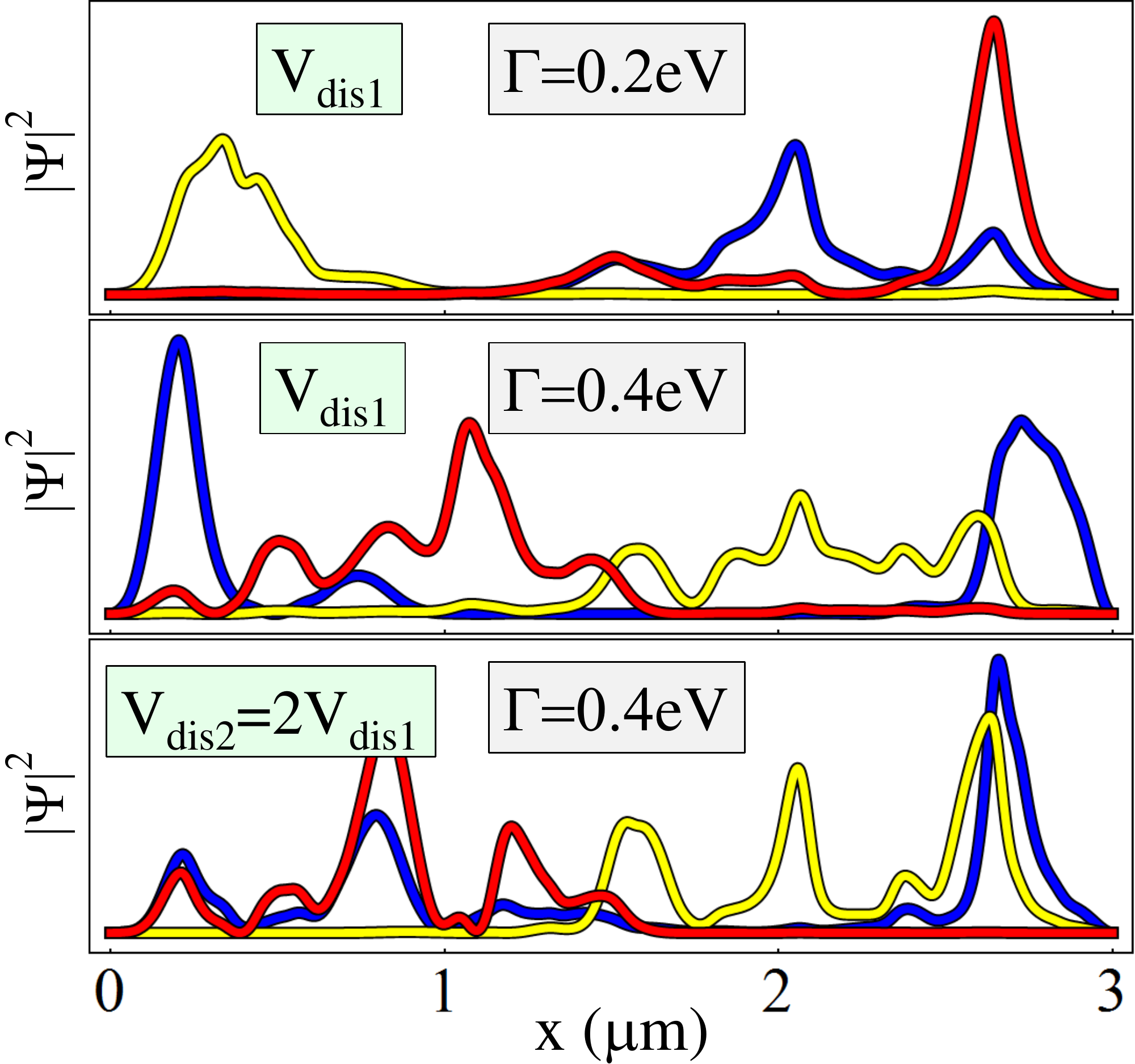}
\vspace{-10mm}
\end{center}
\caption{(Color online)  Spatial profiles of the three lowest energy states in a disordered nanowire. The blue (dark gray) lines correspond to the lowest energy state. Note that in the topologically trivial phase (top panel) the lowest energy state is localized in some region inside the wire, while for a Zeeman field corresponding to the topological SC phase (middle) there are two clear peaks at the ends of the wire corresponding to the Majorana bound states. Increasing the strength of the disorder potential  (bottom) results in effectively breaking the wire into smaller segments with Majorana bound states localized at their ends.}
\vspace{-4mm}
\label{sFig5}
\end{figure}

If we are interested in the relative strength of various features in the DOS and LDOS,  it is rather difficult to obtain a good estimate based solely on the color scale diagrams. Instead, it is more convenient to calculate $\rho(\omega)$ and $\widetilde{\rho}_{\rm end}(\omega)$  trace at different magnetic fields.  For completeness, we show these traces for three relevant parameter regimes: a) the single occupancy, ``delocalized'' regime is shown in Fig. \ref{sFig2} [to be compared with color scale plot shown in Fig. \ref{sFig1}], b) the multi-band, ``delocalized'' regime is shown in Fig. \ref{sFig3} [to be compared with  Fig. 1 from the main text], and c) the multi-band, ``localized'' regime is shown in Fig. \ref{sFig4} [to be compared with  Fig. 4 from the main text].

We point out that the LDOS is closely related to the experimentally measurable tunneling conductance, but  not necessarily proportional to it in all situations.  However, it is difficult to think of cases where the LDOS manifests no contribution from low energy states (i.e., it is characterized by  gap), while the tunneling conductance reflects the presence of certain in--gap features. As such, our main qualitative finding in this work, i.e. that the Majorana mode may show up in the experiment without the obvious closure of the SC gap, remains valid independent of whether the theory focuses on the LDOS or the tunneling conductance.

{\em Role of disorder}. Next, we show that the presence of disorder in a system with parameters corresponding to the delocalized regime does not modify our conclusion. In other words, if the gap closing signature is strongly suppressed in the end--of--wire LDOS of a clean wire, adding disorder will not generate such a signature. The reason for this  behavior stems from the properties of the low--energy states in the presence of a disorder potential~\cite{SLDS}. For a typical disorder realization, the low--energy states that are extended in a clean system become localized within random segments of the nanowire. Nonetheless, localization near the end of the wire is accidental and, typically, the contribution of these states to the end--of--wire LDOS remains negligible compared to the contribution of the Majorana bound state or the contributions from states associated with low--energy bands, which have large amplitudes near the end of the wire in both clean and disordered wires. To illustrate this behavior, we consider a wire of length $L_x=3\mu$m and a chemical potential at the bottom of the forth band ($\Delta\mu=0$) in the presence of a random potential $V_{\rm dis}({\bm i})$ with a characteristic length scale~\cite{SLDS} of the order of the wire width, $L_y$. The dependence of the amplitude of the three lowest energy states on the position along the wire is shown in Fig. \ref{sFig5}.  Note that the amplitude of the low-energy states go smoothly to zero at the ends of the wire, except the Majorana bounds state (blue line in the middle panel) that is characterized by a sharp rise.      

\begin{figure}[tbp]
\begin{center}
\includegraphics[width=0.48\textwidth]{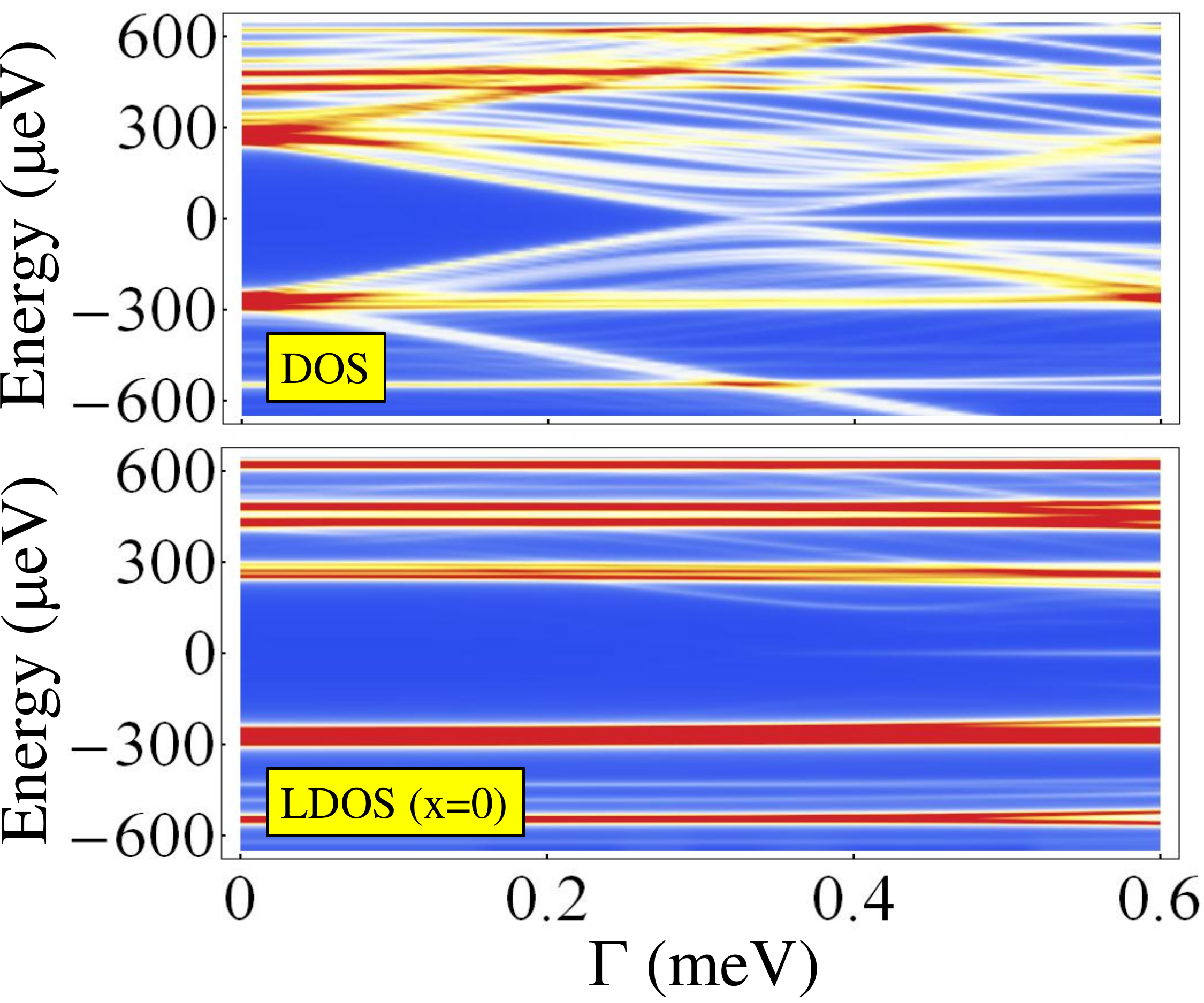}
\vspace{-10mm}
\end{center}
\caption{(Color online)  Dependence of the density of states (top panel) and the end--of--wire LDOS (bottom) on the Zeeman field for a disordered wire with the same parameters as in Fig. \ref{sFig5} ($L_x=3\mu$m, $\Delta\mu=0$, and $V_{\rm dis}=V_{\rm dis1}$). Similar to the clean case, the closing of the quasiparticle gap is clearly seen the the total DOS, but it is strongly suppressed in the LDOS. The weakly Zeeman field--dependent  features that are dominant in the LDOS are due to states from the lower--energy bands that have a large amplitude at the end of the wire.}
\vspace{-4mm}
\label{sFig6}
\end{figure}

\begin{figure}[tbp]
\begin{center}
\includegraphics[width=0.48\textwidth]{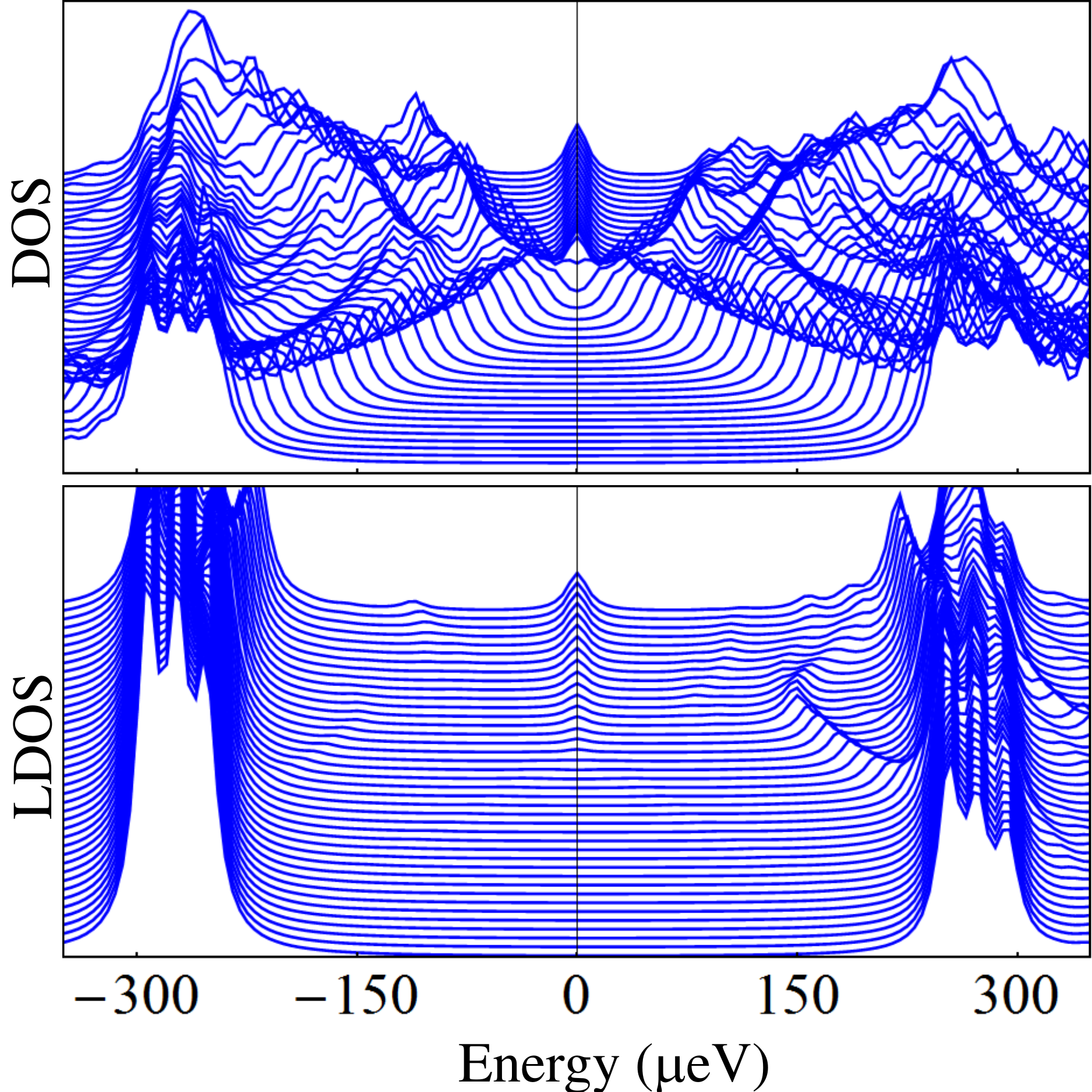}
\vspace{-10mm}
\end{center}
\caption{(Color online)  Zeeman field traces for the DOS (top) and the LDOS (bottom) shown in Fig. \ref{sFig6}. The traces are offset for clarity. Due to the presence of disorder, a few states with energy lower than the induced gap $\Delta_{\rm ind}=250\mu$eV are localized near the end of the wire and generate features in the LDOS. However,  similar to the clean case, there is no gap--closing signature.}
\vspace{-4mm}
\label{sFig7}
\end{figure}

The spatial properties of the low--energy states translate directly into the dependence of the end--of--wire LDOS on the Zeeman field (see Fig. \ref{sFig6} and Fig. \ref{sFig7}). 
Similar to the clean case, the strongest contributions to the end--of--wire LDOS come from states from the lower energy bands that disperse weakly with the Zeeman field. If a low-energy state from the Majorana band (the top occupied band) is accidentally localized near the end of the wire, it will will generate a contribution to the end--of--wire LDOS that disperses with the magnetic field. Such contributions can be identified in Figures \ref{sFig6} and \ref{sFig7}. However, in order to have a clear gap closing signature (e.g., similar to that shown in the lower panel of Fig. \ref{sFig4}), it is necessary that the lowest energy mode be localized near the end of the wire in the topologically trivial case. This requirement is not satisfied for a typical disorder realization. Hence, for a system in  the ``delocalized'' regime, the end--of--wire LDOS will be characterized by the absence of a gap closing signature, even in the presence of disorder. 

{\em Role of finite temperature}.  Finite temperature does not affect our conclusion adversely at all, in fact, reinforcing our finding.  More specifically, finite temperature produces broadening effects, but does not shift spectral weight. Consequently, any feature in the measured tunneling conductance that could be associated with the closing of the gap is sharpest at $T=0$, while finite temperature reduces or suppresses it. Nonetheless, if the weight of a gap--closing feature is comparable to that of the Majorana ZBCP, it cannot be completely suppressed by temperature broadening without destroying  the ZBCP itself. Furthermore, a feature that is not present at $T=0$ is unlikely to appear at finite temperature. Thus, finite temperature cannot provide a mechanism for the non--closure of the gap at the topological quantum phase transition, which is the mystery being addressed in our work.  It should be emphasized that the actual temperature regime used in the experiments is substantially below the SC gap value, hence it has a very small quantitative impact on the results presented in this work.  

\end{document}